\documentclass[aps,pra,twocolumn,amsmath,amssymb,showpacs,nofootinbib,superscriptaddress]{revtex4}

\newcommand{\bra}[1]{\langle#1|}
\newcommand{\ket}[1]{|#1\rangle}

\usepackage[dvips]{graphicx}
\usepackage{mathrsfs}

\begin{document}

\bibliographystyle{apsrev}

\title{Spectral structure and decompositions of optical states, and their applications}

\author{Peter P. Rohde} \email[]{rohde@physics.uq.edu.au}
\homepage{http://www.physics.uq.edu.au/people/rohde/}
\affiliation{Centre for Quantum Computer Technology, Department of Physics\\ University of Queensland, Brisbane, QLD 4072, Australia}
\affiliation{Max Planck Research Group, G{\" u}nther-Scharowsky-Str. 1 / Bau 24, 91058 Erlangen, Germany}

\author{Wolfgang Mauerer} \author{Christine Silberhorn}
\affiliation{Max Planck Research Group, G{\" u}nther-Scharowsky-Str. 1 / Bau 24, 91058 Erlangen, Germany}

\date{\today}

\frenchspacing

\begin{abstract}
We discuss the spectral structure and decomposition of multi-photon states. Ordinarily `multi-photon states' and `Fock states' are regarded as synonymous. However, when the spectral degrees of freedom are included this is not the case, and the class of `multi-photon' states is much broader than the class of `Fock' states. We discuss the criteria for a state to be considered a Fock state. We then address the decomposition of general multi-photon states into bases of orthogonal eigenmodes, building on existing multi-mode theory, and introduce an occupation number representation that provides an elegant description of such states. This representation allows us to work in bases imposed by experimental constraints, simplifying calculations in many situations. Finally we apply this technique to several example situations, which are highly relevant for state of the art experiments. These include Hong-Ou-Mandel interference, spectral filtering, finite bandwidth photo-detection, homodyne detection and the conditional preparation of Schr\"odinger Kitten and Fock states. Our techniques allow for very simple descriptions of each of these examples.
\end{abstract}


\maketitle

\section{Introduction}
Quantum communication and optical quantum computation rely on the controlled preparation and manipulation of specific, well-defined photonic states. In recent decades two main frameworks for the implementation of quantum networks have been established: for systems based on single-photon states the quantum information is encoded in  mode properties (e.g. polarization, frequency or optical phases) of the optical fields, while it is normally assumed that each channel contains exactly one photon. Contrariwise, multi-photon states are typically employed in the context of continuous variable (CV) coding where the quantum observables are conjugate quadratures which are directly linked to the photon numbers of the fields. In this case the description of the field properties is mostly reduced  to a single-mode phase space representation \cite{bib:Vogel06} ignoring spatio-spectral degrees of freedom.

For single-photon states the impact of the spatio-spectral structure has been the subject of previous investigation \cite{bib:URen03,bib:Law00,bib:Law04,bib:Legero03}. Experimentally there exist different approaches to realize single-photon states. If the emission of single quantum systems is utilized, e.g. by employing single atoms, molecules or quantum dot systems (for review see e.g. \cite{bib:Oxborrow05}), the suppression of any higher photon number states for each individual creation event avoids a complex internal structure. As a common alternative conditional single-photon states can be generated by spontaneous parametric downconversion, which produces distinguishable photon pairs. In this case the detection of a trigger photon heralds the existance of a signal photon, but the spatial and spectral correlations between the photon pairs have a large impact on the state purity and thus on performance of quantum interferences in networks \cite{bib:URen05}. Different methods for engineering the spectral properties of spontaneous parametric down-conversion processes for conditional single-photon preparation \cite{bib:Grice01,bib:Branning99} and entanglement based applications have been studied \cite{bib:Carrasco06,bib:Grice97}. For bi-photon states the Schmidt decomposition has been proven to be a successful tool to analyze the properties the spatial and spectral degrees of freedom of quantum fields \cite{bib:URen05,bib:HuangEberly93,bib:ParkerBosePlenio00,bib:Law00,bib:Lvovsky06}. The multi-mode character of multi-photon states is often ignored or may be irrelevant. Though for the generation of pulsed multi-photon squeezed states, spectral properties and inter-mode correlations play an important role in squeezing optimization and for the complete characterization of realistic sources \cite{bib:Wasilewski06,bib:Opatrny02}. Recent experiments now employ for the first time squeezed states in combination with conditional state preparation to realize non-Gaussian quantum states with negative Wigner functions \cite{bib:Ourjoumtsev06,bib:Neergard-Nielsen06,bib:Lvovsky01}. These are crucial for further progress in CV quantum communication \cite{bib:Browne03} and computation \cite{bib:Menicucci06} applications, but the theoretical modelling so far is frequently restricted to a single-mode representation.

The  need for more complex multi-photon states of light makes the accurate description of such states in an elegant and compact representation an important issue. Most notably, in the context of quantum networks and conditional state preparation it is important to realize that different physical processes can \emph{impose} a particular basis choice, because they operate on only a particular well-defined mode. For example, filtering and detection processes are characterized by a particular spectral-temporal response. In this paper we analyze the internal spectral structure of multi-photon states and study its implications on quantum interference effects. We discuss the properties, symmetry and decomposition of the spectral distribution functions of generalized states with higher photon number components. Additionally, we present techniques for performing multi-mode calculations with such states. Previous work has introduced a multi-mode theory \cite{bib:Banaszek02} for the specific case of homodyne detection, where an optical field is decomposed into components orthogonal to and overlapping with some desired mode. We expand upon this notion with a generalized eigenmode decomposition for arbitrary states.
Our theoretical model provides us with great freedom in choosing the basis of the decomposition,  while remaining completely general. The flexibility in choosing the eigenmode decomposition basis makes it very suitable for describing many physical systems of experimental interest. In quantum networks the physical processes impose a decomposition in terms of the following basis: One mode is defined by the response, and all others are moved to a different Hilbert space and may be traced out. Thus, the response of the filtering/detection can significantly change the nature of the state. This can have significant implications on, for example, conditional state preparation techniques, or detected characteristics linked to photon-number statistics, e.g.  phase space representations reconstructed from homodyne measurements. Our approach reduces the complexity of calculations which involve higher photon number states and/or multiple modes and highlights the influence of the internal structure of pulsed photonic states. In particular for multi-photon statesour analysis indicates the limitations of single-mode description where the frequencies and wavevectors of the involved fields are not taken into account.

We begin by introducing a general representation for optical states, which includes both the photon number and spectral degrees of freedom in Section~\ref{sec:gen_rep}. We review the structure and normalization of such states and demonstrate that these states are far more general than the usual notion of `Fock' states. We then analyze the relationship between generalized multi-photon states and Fock states, and give criteria for photonic states to be formally regarded as Fock states. We describe the impact on quantum interference effects. In Section~\ref{sec:eigen_decomp} we introduce a decomposition of states into discrete, orthogonal `eigenmodes', as
well as an `eigenmode occupation number' representation. This provides an elegant representation for general states, and is a useful tool when performing calculations. For example, if we wish to understand a specific interaction between two arbitrary states, we proceed to decompose them into matched discrete eigenmodes. The overall evolution is then given by independently applying the relevant interaction to each eigenmode. We thus take advantage of the fact that photons within a given eigenmode are by definition indistinguishable and can be treated in the idealized sense. In Section~\ref{sec:examples} we demonstrate the benefits of our technique by applying it to several examples of experimental relevance. The described decomposition techniques allow for very straightforward analysis of these model situations. While we treat the detection in a rather idealized way, it becomes apparent that already the generation of multi-photon states with internal  spatio-spectral correlations prohibit a straight forward single-mode discription in many cases. Narrowband filtering can only be applied in particular experiments to improve the purity of the quantum states.
 However, our techniques could also prove very useful in the theoretical analysis of other quantum optical systems, in particular those subject to non-ideal effects such as mode-mismatch \cite{bib:RohdeRalph05,bib:RohdePryde05,bib:RohdeRalphMunro06,bib:RohdeRalph06}, photon distinguishability and imperfect photo-detection \cite{bib:RohdeRalph06b}. We conclude in Section~\ref{sec:conclusion}.

\section{Generalized representation of multi-photon states} \label{sec:gen_rep}

\subsection{Theory}
We begin by discussing the representation of optical states in the spectral domain. Note that while we focus on the spectral degrees of freedom, the discussed representations can easily be generalized to other degrees of freedom, such as the transverse spatial ones.

Including the spectral degrees of freedom, a pure $n$-photon state can be expressed in the form
\begin{equation} \label{eq:spectral_n_photon}
\ket{\psi_n} \propto \int \psi(\omega_1,\dots,\omega_n) \hat{a}^\dag(\omega_1) \dots \hat{a}^\dag(\omega_n) \, \mathrm{d}\vec\omega \ket{0},
\end{equation}
where $\hat{a}^\dag(\omega)$ is the single frequency photonic creation operator at frequency $\omega$. $\psi$ is a continuous function which characterizes the spectral distribution of the state. We refer to this as the spectral distribution function (SDF). An arbitrary pure state can then be expressed in the form
\begin{equation}
\ket{\psi} \propto \sum_n c_n'\int \psi_n(\omega_1,\dots,\omega_n) \hat{a}^\dag(\omega_1) \dots \hat{a}^\dag(\omega_n) \, \mathrm{d}\vec\omega\ket{0}.
\end{equation}

We now turn our attention to the normalization of such states. Our discussion closely follows the work of Ou \cite{bib:Ou06}. Let us assume the SDF is initially normalized according to
\begin{equation} \label{eq:sep_norm}
\int \left| \psi(\omega_1,\dots,\omega_n) \right |^2\mathrm{d}\vec\omega = 1.
\end{equation}
For a single photon state this is equivalent to requiring that the state itself be normalized, i.e. $\langle\psi|\psi\rangle = 1$. However, for states with higher photon number this is not necessarily the case (i.e. in general $\langle\psi|\psi\rangle\neq1$), as we now discuss. For the $n$-photon state $\ket{\psi_n}$, we introduce the normalization term, $\mathcal{N}_n$, defined such that $\langle\psi_n|\psi_n\rangle=1$. Thus, a normalized multi-photon state is of the form
\begin{equation}
\ket{\psi} = \sum_n \frac{c_n}{\sqrt{\mathcal{N}_n}} \int \psi_n(\omega_1,\dots \omega_n) \hat{a}^\dag(\omega_1) \dots \hat{a}^\dag(\omega_n) \, \mathrm{d}\vec\omega\ket{0},
\end{equation}
where $\sum_n |c_n|^2=1$. Following Ref.~\cite{bib:Ou06}, the normalization terms are given by
\begin{equation} \label{eq:N_def}
\mathcal{N}_n = \int \psi_n(\omega_1,\dots,\omega_n)^* \sum_{P\in S_n} \psi_n(P[\omega_1,\dots,\omega_n]) \,\mathrm{d}\vec\omega,
\end{equation}
where $P$ represents a permutation of the indices of $\psi$, and we sum over all possible permutations. This expression arises directly by expanding $\langle\psi_n|\psi_n\rangle$. This expansion consists of $n!$ terms, containing all possible combinations of $n$-fold products of delta functions of the form $\langle \omega_i | \omega_j \rangle=\delta(\omega_i-\omega_j)\,\,\forall\,\,i,j=1\dots n$. This gives rise to the summation over permutations of the indices. Since there are $n!$ terms in the summation, the factor $\mathcal{N}_n$ ranges between 1 and $n!$, and directly gives us a measure of the permutation symmetry of $\psi$.

To provide some intuition into the behavior of $\mathcal{N}$, let us consider two simple examples. First, suppose the SDF can be expressed in the form
\begin{equation} \label{eq:sep_indis_sdf}
\psi(\omega_1,\dots,\omega_n) = \phi(\omega_1)\dots\phi(\omega_n).
\end{equation}
In this case $\psi$ exhibits full symmetry under permutation of the indices, i.e. $\psi(P[\omega_1,\dots,\omega_n])=\psi(\omega_1,\dots,\omega_n)\,\,\forall\,\,P\in S_n$, and it follows from Eq.~\ref{eq:N_def} that $\mathcal{N}=n!$. It is shown in Appendix~\ref{app:n_fact_proof} that $\mathcal{N}=n!$ if and only if $\psi$ exhibits full permutation symmetry. Note that full permutation symmetry does not uniquely correspond to the class of states of the form shown in Eq.~\ref{eq:sep_indis_sdf}, i.e. indistinguishable separable states. For example, if we take the SDF to be a linear combination of two distinct such functions, e.g. $\psi(\omega_1,\omega_1)=\alpha\phi(\omega_1)\phi(\omega_2) + \beta\varphi(\omega_1)\varphi(\omega_2)$, it will also exhibit full
permutation symmetry, but is non-separable.

Next, suppose the multi-photon state consists of temporally-separated photons, sufficiently separated such that every photon is distinguishable from the others. This essentially corresponds to $n$ photons in $n$ distinct modes, as opposed to $n$ photons in one mode as per the previous example. Now the SDF can be expressed in the form
\begin{eqnarray} \label{eq:sep_dis_sdf}
\psi(\omega_1,\dots,\omega_n) &=& e^{-i\omega_1\tau} \phi(\omega_1)e^{-i2\omega_2\tau} \dots e^{-ni\omega_n\tau} \phi(\omega_n)\nonumber\\
&=& \prod_{k=1}^n e^{-ki\omega_k\tau} \phi(\omega_k)=\prod_{k=1}^n \varphi_k(\omega_k),
\end{eqnarray}
where $\tau$ is the temporal separation between photons and is larger than the temporal bandwidth of the individual photons, i.e. $\tau\gg 1/\Delta\omega$. Thus,
\begin{equation}
\int \varphi_i(\omega)^*\varphi_j(\omega)\approx\delta_{i,j}.
\end{equation}
In this case it is obvious that the state exhibits no permutation symmetry, since the state decomposes such that each photon effectively has zero overlap with all others. Thus, the only surviving term in the expansion is for $P=\openone$, giving $\mathcal{N}=1$.

Note that in the previous two examples the $\mathcal{N}$ normalization factor is consistent with what we expect from standard theory. In the former case we have $n$ excitations of a single mode, in which case standard theory tells us that $\ket\psi = \frac{1}{\sqrt{n!}}(\hat{a}^\dag)^n\ket{0}$. On the other hand, in the latter case we effectively have $n$ distinct modes, each with a single excitation, for which $\ket\psi = \hat{a}_1^\dag \hat{a}_2^\dag\dots\hat{a}_n^\dag\ket{0}$.

This notion generalizes very simply to the multi-mode case. Consider the $m$-partite system with $n_i$ photons in the $i^\mathrm{th}$ mode. Such a state can be expressed in the form
\begin{eqnarray}
&&\ket{\psi_{n_1,\dots,n_m}} = \int \psi(\omega_1^{(1)},\dots,\omega_{n_1}^{(1)},\dots,\omega_1^{(m)},\dots,\omega_{n_m}^{(m)}) \nonumber\\
&&\times\hat{a}^\dag_1(\omega_1^{(1)})\dots\hat{a}^\dag_1(\omega_{n_1}^{(1)}) \dots\hat{a}^\dag_m(\omega_1^{(m)})\dots\hat{a}^\dag_m(\omega_{n_m}^{(m)}) \,\mathrm{d}\vec\omega\ket{0}.\nonumber\\
\end{eqnarray}
where $\omega_i^{(j)}$ denotes the frequency of the $i^\mathrm{th}$ photon in mode $j$, and $\hat{a}^\dag_i(\omega)$ denotes the single frequency creation operator of mode $i$. In this case it follows that the normalization factor $\mathcal{N}$ is given by
\begin{equation}
\mathcal{N}_{n_1,\dots,n_m}=\int \psi(\vec\omega)^* \sum_{P_1,\dots,P_m} \psi(P_1\dots P_m[\vec\omega]) \,\mathrm{d}\vec\omega,
\end{equation}
where $P_i$ is a permutation over the just the indices of mode $i$, and again we sum over all permutations. It can easily be seen that in the multimode case $\mathcal{N}$ varies between $1$ and $n_1!n_2!\dots
n_m!$.

\subsection{What is a Fock state?} \label{sec:fock_vs_not}
It is evident that there is an entire class of states consisting of $n$ photons. This raises the question as to what subset of these states is equivalent to the usual notion of Fock  states. We answer this question by examining the algebraic structure of Fock states. Formally, the $n$ photon Fock state is defined as
\begin{equation}
\ket{n}=\frac{1}{\sqrt{n!}}\left(\hat{a}^\dag\right)^n\ket{0}.
\end{equation}
To maintain a consistent algebraic structure we require that states defined in the more generalized spectral representation exhibit an analogous structure,
\begin{equation} \label{eq:fock_spec}
\ket{n}=\frac{1}{\sqrt{n!}}\left(\hat{A}_\mathrm{\psi}^\dag\right)^n\ket{0},
\end{equation}
where $\hat{A}_\mathrm{\psi}^\dag$ is a generalized mode creation operator, which creates a photon characterized by SDF $\psi(\omega)$,
\begin{equation}
\hat{A}_{\mathrm{\psi}}^\dag = \int \psi(\omega)\hat{a}^\dag(\omega) \, \mathrm{d}\omega.
\end{equation}
It is clear that the class of states corresponding to the Fock states is defined by the constraint on the multi-photon SDF,
\begin{equation}
\psi(\omega_1,\dots,\omega_n)=\phi(\omega_1)\phi(\omega_2)\dots\phi(\omega_n),
\end{equation}
where $\phi(\omega)$ is arbitrary. Thus, when the SDF of an $n$-photon state is factorizable as an $n$-fold product of identical functions, it is structurally equivalent to a Fock state. Henceforth we will refer to states of this form as Fock states. Note that for Fock states $\mathcal{N}=n!$, giving rise to the required normalization factor of $1/\sqrt{n!}$ in Eq.~\ref{eq:fock_spec}. Importantly, although Fock states necessarily satisfy $\mathcal{N}=n!$, states satisfying $\mathcal{N}=n!$ are not necessarily Fock states. An example of this are entangled states of the form $\psi(\omega_1,\omega_1)=\alpha\phi(\omega_1)\phi(\omega_2) + \beta\varphi(\omega_1)\varphi(\omega_2)$, that were discussed earlier. Such states are fully permutation symmetric but are not Fock states.

Often spectral distinguishability of multi-photon states is ignored, and an $n$-photon Fock state and a state with $n$ photons are regarded as being equivalent. As we have discussed, Fock states only correspond to a subset of the later more general class of states. It is not immediately obvious what implications this has in an experimental
situation. An example of this question was studied in Ref.~\cite{bib:Ou06}, and temporal distinguishability in multi-photon states has been experimentally demonstrated in Ref.~\cite{bib:Xiang06}.

To illustrate this, and provide some intuition into the importance of the distinction between Fock states and more general multi-photon states, we now discuss a very simple example situation where the behavior of an experiment differs significantly, depending on the spectral composition of multi-photon states.

Consider an experiment where we interfere two identical copies of a two-photon state on a 50/50 beamsplitter. We will consider two opposing limits. First, we employ Fock states, as defined in the previous discussion. In this case each of the incident two-photon states are of the spectral form $\psi(\omega_1,\omega_2)=\phi(\omega_1)\phi(\omega_2)$. Second, we employ two-photon states in which the constituent photons are well temporally separated. In this case,
$\psi(\omega_1,\omega_2)=e^{-i\omega_1\tau}\phi(\omega_1)\phi(\omega_2)$, where $\tau$ is much larger than the photons' temporal bandwidth.

Let us consider the probability of detecting four photons at output mode $A$, $P_{4A}$. In the former case the derivation proceeds in the usual manner.
\begin{eqnarray}
\ket{\psi_\mathrm{in}} &=& \frac{1}{2} (\hat{a}^\dag)^2 (\hat{b}^\dag)^2\ket{0} \nonumber\\
\ket{\psi_\mathrm{out}} &=& \hat{U} \ket{\psi_\mathrm{in}} \nonumber\\
&=& \left[ \frac{(\hat{a}^\dag)^4}{8} - \frac{(\hat{a}^\dag)^2 (\hat{b}^\dag)^2}{4} + \frac{(\hat{b}^\dag)^4}{8}\right]\ket{0}\nonumber\\
&=&\sqrt{\frac{3}{8}}\ket{4,0}-\frac{1}{2}\ket{2,2}+\sqrt{\frac{3}{8}}\ket{0,4},
\end{eqnarray}
where $\hat{U}$ is the 50/50 beamsplitter operation and we assume the phase-asymmetric beamsplitter convention. Therefore $P_{4A}=3/8$.

In the later case, temporally disjoint photons do not interact with one another at the beamsplitter. We can therefore treat the evolution as two independent interferences of single photons at the beamsplitter. For a single such event we have
\begin{eqnarray}
\ket{\psi_\mathrm{in}} &=& \hat{a}^\dag \hat{b}^\dag\ket{0} \nonumber\\
\ket{\psi_\mathrm{out}} &=& \hat{U}\ket{\psi_\mathrm{in}} \nonumber\\
&=&\left[\frac{(\hat{a}^\dag)^2}{2} - \frac{(\hat{b}^\dag)^2}{2}\right]\ket{0} \nonumber\\
&=& \frac{1}{\sqrt{2}} \ket{2,0} - \frac{1}{\sqrt{2}} \ket{0,2}.
\end{eqnarray}
Thus, for each event the probability of detecting two photons at any one of the outputs is $1/2$, as expected from the well known Hong-Ou-Mandel (HOM) effect \cite{bib:HOM87}. Given two independent events of this type, the total probability of detecting four photons at output $A$ is $P_{4A}=1/4$.

Although in both cases we are interacting two indistinguishable states, it is evident that the nature of the quantum interference differs significantly depending on their internal structure.

These two limiting cases are easily calculated from standard theory. However, we lack a description of the more general case where the two-photon states exhibit partial internal distinguishability. We now consider a more general case. Suppose our incident two-photon states are separable and of the form
\begin{equation}
\ket{\psi_2} = \frac{1}{\sqrt{\mathcal{N}_2}}\hat{A}_{\varphi_1}^\dag\hat{A}_{\varphi_2}^\dag\ket{0}.
\end{equation}
where $\varphi_1(\omega)$ and $\varphi_2(\omega)$ are arbitrary and may exhibit any degree of distinguishability. It can easily be seen that the normalization factor $\mathcal{N}_2$ is given by
\begin{eqnarray}
\mathcal{N}_2 &=& \left(\int \varphi_1(\omega)^*\varphi_1(\omega)\,\mathrm{d}\omega\right) \left(\int \varphi_2(\omega)^*\varphi_2(\omega)\,\mathrm{d}\omega\right)\nonumber\\
&+& \left(\int \varphi_1(\omega)^*\varphi_2(\omega)\,\mathrm{d}\omega\right) \left(\int \varphi_2(\omega)^*\varphi_1(\omega)\,\mathrm{d}\omega\right) \nonumber\\
&=& 1+\gamma,
\end{eqnarray}
where
\begin{equation}
\gamma = \left|\int \varphi_1(\omega)^* \varphi_2(\omega)\,\mathrm{d}\omega\right|^2
\end{equation}
characterizes the degree of photon distinguishability within the two-photon state. We have two identical copies of this state, so the incident state is of the form
\begin{equation}
\ket{\psi_\mathrm{in}} = \frac{1}{\mathcal{N}_2} \hat{A}_{\varphi_1}^\dag \hat{A}_{\varphi_2}^\dag \hat{B}_{\varphi_1}^\dag \hat{B}_{\varphi_2}^\dag \ket{0},
\end{equation}
where $A$ and $B$ denote distinct spatial modes. We then pass this state through a 50/50 beamsplitter and condition upon detecting all four photons at the $A$ output. When applying the beamsplitter transformation we assume that corresponding spectral modes from spatial modes $A$ and $B$, match one another at the beamsplitter (we make this assumption throughout this paper). It can now easily be seen that the conditional output state is
\begin{equation}
\ket{\psi_\mathrm{cond}} = \frac{1}{4\mathcal{N}_2}\hat{A}_{\varphi_1}^\dag \hat{A}_{\varphi_1}^\dag \hat{A}_{\varphi_2}^\dag \hat{A}_{\varphi_2}^\dag \ket{0}.
\end{equation}
The probability of detecting this state is simply given by $P_{4A}=\langle\psi_\mathrm{cond}|\psi_\mathrm{cond}\rangle$. Thus,
\begin{equation}
P_{4A} = \frac{\mathcal{N}_4}{16{\mathcal{N}_2}^2},
\end{equation}
where $\mathcal{N}_4$ is the normalization factor of the four-photon state $\hat{A}_{\varphi_1}^\dag \hat{A}_{\varphi_1}^\dag \hat{A}_{\varphi_2}^\dag \hat{A}_{\varphi_2}^\dag \ket{0}$. Notice that the interference is completely characterized by the relationship between different permutation symmetry factors. It can easily be shown
(see Appendix~\ref{app:N_4_proof} for proof) that in this case $\mathcal{N}_4 = 4(1+\gamma+\gamma^2)$. Thus we have
\begin{equation}
P_{4A} = \frac{1}{4}\left(\gamma+\frac{1}{1+\gamma}\right).
\end{equation}
As expected, this is consistent with the limiting cases considered previously. When we have incident Fock states, $\gamma=1$ and $P_{4A}=3/8$, whereas for states in which the constituent photons are distinguishable we have $\gamma=0$ and $P_{4A}=1/4$.

\section{Eigenmode decomposition of multi-photon states} \label{sec:eigen_decomp}

\subsection{Motivation}
From the previous section it is evident that the internal spectral structure of multi-photon states plays an important role in quantum interference effects. The calculation of the previous example also illustrates that it becomes increasingly difficult to model generalized settings with higher photon numbers, since we must
analytically manipulate complicated multi-fold integrals with increasing numbers of terms. In this section we discuss an alternative approach for representing multi-photon states in a discrete basis of eigenfunctions, which provides an elegant and compact description for such states and simplifies many calculations. We will demonstrate its benefits in the next section by analyzing several example situations. Our method lends itself well to numerical analysis, as opposed to analytic manipulation of states in integral form. This approach is an extension of the multi-mode description of optical states, which has been used in the evaluation of homodyne detection \cite{bib:Banaszek02}, and is also closely related to the well-known Schmidt decomposition for two-mode states.

\subsection{Theory}
Any well-behaved complex function (in the present discussion a SDF) can always be decomposed in terms of a discrete basis of orthonormal functions (a well known example is the basis of Hermite functions),
\begin{equation}
\psi(\omega) = \sum_i \lambda_i \xi_i(\omega).
\end{equation}
where $\sum_i |\lambda_i|^2=1$, and the discrete set of functions $\{\xi_i(\omega)\}$ are orthonormal under the inner product,
\begin{equation}
\int \xi_i(\omega)^* \xi_j(\omega) \,\mathrm{d}\omega = \delta_{i,j}.
\end{equation}
In the present discussion the exact choice of eigenfunctions will typically be unimportant. However, we will extensively make use of one property: with an appropriate change of basis transformation we can choose an arbitrary normalizable function to belong to the discrete set of basis functions. This derives from the Gram-Schmidt orthogonalization procedure and is proven in Appendix~\ref{app:function_decomp_proof}.

It follows that a single photon state with arbitrary SDF can always be expressed as a superposition of single photon states with orthogonal SDF's,
\begin{eqnarray}
\ket{\psi_1} = \int \psi(\omega)\hat{a}^\dag(\omega)\,\mathrm{d}\omega\ket{0} &=& \sum_i \lambda_i \int \xi_i(\omega)\hat{a}^\dag(\omega)\,\mathrm{d}\omega\ket{0}\nonumber\\
&=& \sum_i \lambda_i \hat{A}^\dag_{\xi_i}\ket{0}.
\end{eqnarray}
Because the basis functions $\{\xi_i(\omega)\}$ are orthonormal, $\bra{0}\hat{A}_{\xi_i}\hat{A}^\dag_{\xi_j}\ket{0}=\delta_{i,j}$. Thus, the modes characterized by the different mode creation operators $\hat{A}^\dag_{\xi_i}$ are orthonormal. We refer to these as `eigenmodes'. The coefficients $\{\lambda_i\}$ follow trivially from the overlap $\lambda_i = \bra{0}\hat{A}_{\xi_i}\hat{A}_\psi^\dag\ket{0}$,
\begin{equation}
\lambda_i = \int \xi_i(\omega)^*\psi(\omega)\,\mathrm{d}\omega.
\end{equation}
This notion generalizes to states with higher photon number in a straightforward manner,
\begin{eqnarray}
\ket{\psi_n} &=& \int \psi(\omega_1,\dots,\omega_n)\hat{a}^\dag(\omega_1)\dots\hat{a}^\dag(\omega_n)\,\mathrm{d}\vec\omega\ket{0}\nonumber\\
&=& \sum_{i_1\leq\dots\leq i_n} \lambda_{i_1,\dots,i_n} \nonumber\\
&&\times \int \xi_{i_1}(\omega_1)\dots\xi_{i_n}(\omega_n)\hat{a}^\dag(\omega_1)\dots\hat{a}^\dag(\omega_n)\,\mathrm{d}\vec\omega\ket{0}\nonumber\\
&=& \sum_{i_1\leq\dots\leq i_n} \lambda_{i_1,\dots,i_n} \hat{A}^\dag_{\xi_{i_1}}\dots\hat{A}^\dag_{\xi_{i_n}}\ket{0}.
\end{eqnarray}
Notice that we insert the summation condition $i_1\leq\dots\leq i_n$ to remove potential ambiguity in the
decomposition. $\lambda_{i_1,\dots,i_n}$ is to be interpreted as the coefficient of the term with a photon in each of the modes $i_1,\dots,i_n$. Therefore $\lambda$'s with different permutations of the same indices refer to the same eigenmode. For example, in the two photon state the coefficients $\lambda_{1,2}$ and $\lambda_{2,1}$ are both associated with a single photon in each of the modes $\xi_1$ and $\xi_2$. Thus, to avoid this double counting we assume the summation does not run over permutations of the same indices. Now the coefficients of the decomposition are uniquely defined as
\begin{equation} \label{eq:lambda_def}
\lambda_{i_1,\dots,i_n} = \int \xi_{i_1}(\omega_1)^*\dots\xi_{i_n}(\omega_n)^*\psi(\omega_1,\dots,\omega_n)\,\mathrm{d}\vec\omega.
\end{equation}

We can reformalize this representation slightly as follows. Consider the single photon case. We define
$\ket{1}_{\xi_i}=\hat{A}^\dag_{\xi_i}\ket{0}$ to be the single photon state in the eigenmode characterized by $\xi_i$. Because the different eigenmodes are orthonormal, we can introduce an occupation number representation,
\begin{equation}
\ket{1}_{\xi_i} = \ket{0}_{\xi_1}\ket{0}_{\xi_2}\dots\ket{1}_\mathrm{\xi_i}\ket{0}_{\xi_{i+1}}\dots.
\end{equation}
Thus, an arbitrary single photon state can always be expressed in the form
\begin{equation}
\ket{\psi_1}=\sum_i \lambda_i \ket{0}_{\xi_1}\ket{0}_{\xi_2}\dots\ket{1}_\mathrm{\xi_i}\ket{0}_{\xi_{i+1}}\dots.
\end{equation}
Similarly, an arbitrary $n$ photon state can be expressed in the eigenmode occupation number representation as
\begin{eqnarray}
\ket{\psi_n} &=& \sum_{i_1\leq\dots\leq i_n} \sqrt{n_1!n_2!\dots} \lambda_{i_1,\dots,i_n} \ket{n_1}_{\xi_1}\ket{n_2}_{\xi_2}\dots\nonumber\\
&=& \sum_{i_1\leq\dots\leq i_n} \lambda_{i_1,\dots,i_n}' \ket{n_1}_{\xi_1}\ket{n_2}_{\xi_2}\dots,
\end{eqnarray}
where $n_j$ is the number of indices $\{i_1,\dots,i_n\}$ equal to $j$, i.e. the population of the respective eigenmode. This effectively represents an arbitrary state in terms of a superposition of distinct Fock states. This change in notation may seem rather pointless, however this is exactly what we ordinarily do when we use regular Fock
notation -- we abstract away the specific form of the underlying wavefunction to simplify treatment. This notational change turns out to be very useful, as we will see in the examples in the next section.

Next we discuss a slight variation of the eigenmode occupation number representation that simplifies certain types of calculation. This will be particularly useful in some of the examples we discuss shortly. In many calculations we do not explicitly need to consider all of the modes in a given decomposition. Instead, there might be just one mode
which is involved in a given interaction, and others are of no direct interest. In such a situation it would be particularly desirable to decompose a state into just two components: an arbitrary mode that is of interest (e.g. this mode might be involved in a particular interaction), and collectively `other' modes that are not directly
involved in our calculations. For example, suppose we measure a state using a photo-detector that responds to a finite spectral range. By defining one of the eigenmodes to be the spectral response of the detector, we can simply trace out the remaining modes to understand the behavior of the detection process -- the exact form of the rest of
the decomposition is not important. This is one of the examples we will consider in detail shortly.

First consider the single photon case. We represent the single photon SDF as,
\begin{equation}
\psi(\omega) = \lambda_1\phi(\omega) + \lambda_0\bar\phi(\omega),
\end{equation}
where $\phi(\omega)$ represents some `desired' mode, and $\bar\phi(\omega)$ collectively represents the component of
$\psi(\omega)$ that does not overlap with $\phi(\omega)$. Now the mode creation operator can be reexpressed,
\begin{equation}
\hat{A}^\dag_\psi = \lambda_1\hat{A}^\dag_\phi+\lambda_0\hat{A}^\dag_{\bar\phi}.
\end{equation}
where
\begin{eqnarray}
\lambda_1 &=& \int \phi(\omega)^*\psi(\omega)\,\mathrm{d}\omega,\nonumber\\
\lambda_0 &=& \sqrt{1-|\lambda_1|^2}.
\end{eqnarray}
Moving to the occupation number representation this can be written
\begin{equation}
\ket{\psi_1} = \lambda_1\ket{1}_{\phi}\ket{0}_{\bar\phi} + \lambda_0\ket{0}_\phi\ket{1}_{\bar\phi}.
\end{equation}
Of course this can also be easily generalized to states with higher photon number. For example, the decomposition of an arbitrary $n$ photon state in terms of Fock states in mode $\phi$ will take the form,
\begin{equation}
\ket{\psi_n} = \sum_{i=0}^n \lambda_i\ket{i}_\phi\ket{n-i}_{\bar\phi}.
\end{equation}

Thus far we have specifically focussed on the single-mode case. However, these ideas easily generalize to the multi-mode case. Specifically, a state distributed across modes $A,B,C,\dots$, with $n_i$ photons in mode $i$ will take the form,
\begin{eqnarray}
\ket{\psi_{n_A,n_B,\dots}} &=& \sum_{\vec{i}^A,\vec{i}^B,\dots} \lambda_{\vec{i}} \hat{A}_{\xi_{i^A_1}}^\dag \dots \hat{A}_{\xi_{i^A_{n_A}}}^\dag \nonumber\\
&& \times \hat{B}_{\xi_{i^B_1}}^\dag \dots \hat{B}_{\xi_{i^B_{n_B}}}^\dag \times \dots \ket{0}.
\end{eqnarray}
This representation for multi-photon states is closely related to the Schmidt decomposition for biphoton states. Formally, a Schmidt decomposition corresponds to the $n_A=1,\,n_B=1$ case where we choose the eigenbasis $\{\xi_i\}$ so as to diagonalize the $\lambda$ matrix. Unlike the Schmidt decomposition, the representation described is completely general and holds for states with arbitrary photon number.

\section{Experimental Examples} \label{sec:examples}
The eigenmode decomposition and occupation number representation provide an elegant tool for performing calculations with generalized multi-photon states. We now present several examples of experimental relevance, which illustrate the simplicity of modeling such systems using this representation. First we review the well known Hong-Ou-Mandel effect, which provides a simple example scenario for our techniques. Second, we model spectral filtering and finite bandwidth photo-detection, which follows very simply from an occupation number representation, and is important in understanding the behavior of realistic photo-detectors. In the framework of this paper we simplify the calculations by assuming that the detector projects onto a particular, well-defined mode $\ket{\xi_0}$. In practise, realistic photo-detectors may respond to multiple modes but not provide information as to which one is detected. For example, a detector may respond to all times within some window, project onto a particular temporal mode within that window, but not tell us which one. In this instance we will clearly be left with a mixture of states of  different possible detection outcomes. However, if we are interested how internal correlations of multi-photon states impact the general performance an quantum optical system -- derived in a single-mode picture -- an idealized detection provides us with an upper bound for the purity of the states. In the third section
we discus homodyne detection in an effective single-mode description where we introduce the local oscillator mode as the relevant eigenmode.  The fourth and fifth section finally treats two distinct conditional state preparation scenarios: the preparation of Schr\"odinger Kitten states via photon subtraction, and conditional Fock state preparation via non-degenerate parametric down-conversion. These two examples illustrate that the achievalbe purity depends on the internal structure of the initial states as well as on the specific setups.

\subsection{Hong-Ou-Mandel interference}
We begin our discussion by considering the well-known HOM effect \cite{bib:HOM87}. Although this effect is well understood and extremely simple to derive using various other approaches (cf. Ref.~\cite{bib:RohdeRalph05}), for illustration we start by rederiving this result using the techniques discussed. After studying the case of independently produced photons, we  generalize to the case of spectrally entangled photons.

We begin with a two mode input state, with a single photon in each mode. We initially assume the state to be separable, such that the photons are independently characterized by SDFs $\phi(\omega)$ and $\varphi(\omega)$ respectively. The incident state can therefore be expressed in the form
\begin{equation}
\ket\psi_\mathrm{in} = \ket{1}_{\phi,A} \ket{1}_{\varphi,B},
\end{equation}
where $A$ and $B$ denote distinct spatial modes. Shifting to the occupation number representation and choosing one of the eigenmodes to be $\phi(\omega)$ we obtain
\begin{eqnarray}
\ket\psi_\mathrm{in} &=& \ket{1}_{\phi,A}(\lambda_1\ket{1}_{\phi,B}+\lambda_0\ket{1}_{\bar\phi,B})\nonumber\\
&=& \lambda_1\ket{1}_{\phi,A}\ket{1}_{\phi,B}+\lambda_0\ket{1}_{\phi,A}\ket{1}_{\bar\phi,B}.
\end{eqnarray}
Here we have used shorthand notation and not explicitly written out all the
vacuum terms.  We now apply the 50/50 beamsplitter operation
$\hat{U}$. Note that because the different eigenmodes are
orthogonal, and within each eigenmode photons are by definition
indistinguishable, $\hat{U}$ acts on each eigenmode independently
and in the idealized sense. Thus we have the property,
\begin{eqnarray}
&&\hat{U}(\ket{m_1,n_1}_{\xi_1}\ket{m_2,n_2}_{\xi_2}\dots)\nonumber\\
&&=(\hat{U}\ket{m_1,n_1})_{\xi_1}(\hat{U}\ket{m_1,n_2})_{\xi_2}\dots.
\end{eqnarray}
Using this we obtain
\begin{eqnarray}
\ket\psi_\mathrm{out} &=& \hat{U}\ket\psi_\mathrm{in} \nonumber\\
&=& \lambda_1\hat{U}\ket{1,1}_\phi + \lambda_0(\hat{U}\ket{1,0})_\phi(\hat{U}\ket{0,1})_{\bar\phi}\nonumber\\
&=& \frac{\lambda_1}{\sqrt{2}}(\ket{2,0}-\ket{0,2})_\phi \nonumber\\
&+& \frac{\lambda_0}{2}(\ket{1,0}+\ket{0,1})_\phi(\ket{1,0}-\ket{0,1})_{\bar\phi},
\end{eqnarray}
where we have used the shorthand $\ket{m,n}_\phi=\ket{m}_{\phi,A}\ket{n}_{\phi,B}$. From this expression we can directly read off the two-photon coincidence probability $P_c$,
\begin{equation}
P_c = \frac{|\lambda_0|^2}{2} = \frac{1}{2}-\frac{1}{2}\left| \int \phi(\omega)^*\varphi(\omega)\,\mathrm{d}\omega \right|^2.
\end{equation}
This is the usual result -- for completely indistinguishable photons the integral term approaches unity and $P_c=0$, whereas for completely distinguishable photons the integral term approaches zero and $P_c=1/2$.

This analysis effectively describes the situation where we interact
two independently produced photons. Next we consider the more general
case of spectrally entangled photons. This arises, for example, when
using photon pairs produced through spontaneous parametric
down-conversion. By expressing an arbitrary two mode state as
\begin{eqnarray}
  \ket{\psi_\mathrm{in}} &=& \sum_{ij} \lambda_{ij} \hat{A}_i^\dag \hat{B}_j^\dag \ket{0},
\end{eqnarray}
 we can apply the 50/50 beamsplitter operation to obtain
\begin{eqnarray}
  \ket{\psi_\mathrm{out}} &=& \hat{U}\ket{\psi_\mathrm{in}} \nonumber\\
  &=& \frac{1}{2} \sum_{i\neq j} \lambda_{ij} (\hat{A}_i^\dag + \hat{B}_i^\dag)(\hat{A}_j^\dag - \hat{B}_j^\dag) \ket{0} \nonumber\\
  &&+ \frac{1}{2}\sum_i \lambda_{ii} [(\hat{A}_i^\dag)^2 - (\hat{B}_i^\dag)^2] \ket{0}.
\end{eqnarray}
Post-selecting upon coincidence events leaves
\begin{eqnarray}
  \ket{\psi_\mathrm{cond}} &=& \frac{1}{2}\sum_{i\neq j} \lambda_{ij} (\hat{A}_j^\dag \hat{B}_i^\dag - \hat{A}_i^\dag \hat{B}_j^\dag)\ket{0} \nonumber\\
  &=& \frac{1}{2} \sum_{i\neq j} (\lambda_{ij}-\lambda_{ji}) \hat{A}_i^\dag \hat{B}_j^\dag \ket{0}.
\end{eqnarray}
Thus, the coincidence probability is
\begin{equation}
P_c = \frac{1}{4} \sum_{i\neq j} |\lambda_{ij} - \lambda_{ji}|^2.
\end{equation}
We can now immediately understand the behavior of HOM interference in
the general case. First, suppose the input state is separable and the
photons are indistinguishable. Now it is necessarily possible to
choose an eigenbasis such that $\lambda_{11}=1$, otherwise 0. In this
case clearly $P_c=0$, as expected. Next consider the other extreme,
where the two photons are completely distinguishable. Suppose that
photon $A$ resides in eigenmode $\xi_1$ and photon $B$ in eigenmode
$\xi_2$. Now $\lambda_{12}=1$, otherwise 0. Clearly now $P_c=1/2$, as expected. More generally we see that
the requirement for perfect HOM interference is
$|\lambda_{ij}-\lambda_{ji}|=0\,\,\forall\,\,i,j$. This is equivalent
to requiring that the incident two-mode state be exchange
symmetric. In other words, the two-photon joint-SDF
$\psi(\omega_1,\omega_2)$ must be symmetric about the
$\omega_1=\omega_2$ axis. Note that
this is much broader than the class of indistinguishable separable
states, and includes highly spectrally entangled ones. This
explains, for example, why HOM interference works with photon pairs
produced through spontaneous down-conversion, despite the inherent
spectral entanglement in that case (see also Ref.~\cite{bib:Branning00}).

\subsection{Spectral filtering and finite bandwidth photo-detection}
We now consider the situation where we have an $n$-photon state to
which we apply a spectral filter. Suppose the filter transmits
frequencies within some finite range, and completely reflects any
spectral components outside this range. Thus, the filter has an ideal
rectangular spectral response function that we will label
$\phi(\omega)$. For simplicity we will assume the incident state is an
arbitrary Fock state. Note that this technique is by no means limited
to dealing with Fock states, it is just a simplifying assumption for
the sake of example. The incident state can be expressed,
\begin{eqnarray}
  \ket{\psi_n} &=& \frac{1}{\sqrt{n!}} \left(\hat{A}_\psi^\dag\right)^n\ket{0}\nonumber\\
  &=& \frac{1}{\sqrt{n!}} \left(\lambda_1\hat{A}_\phi^\dag + \lambda_0\hat{A}_{\bar\phi}^\dag\right)^n\ket{0}\nonumber\\
  &=& \frac{1}{\sqrt{n!}} \sum_{i=0}^n \binom{n}{i} {\lambda_1}^i {\lambda_0}^{n-i} \left(\hat{A}_\phi^\dag\right)^i \left(\hat{A}_{\bar\phi}^\dag\right)^{n-i} \ket{0}\nonumber\\
  &=& \sum_{i=0}^n \binom{n}{i}^{\frac{1}{2}} {\lambda_1}^i {\lambda_0}^{n-i} \ket{i}_\phi\ket{n-i}_{\bar\phi},
\end{eqnarray}
where we have used an occupation number representation and chosen
one of the eigenmodes to be $\phi(\omega)$. Next, by definition our
filter reflects all modes orthogonal to $\phi(\omega)$. Thus, we trace
them out to obtain the transmitted state,
\begin{equation} \label{eq:rho_filter} \hat\rho = \sum_{i=0}^n
  \binom{n}{i}{\lambda_1}^{2i}{\lambda_0}^{2(n-i)}\ket{i}_\phi
  \bra{i}_\phi,
\end{equation}
which is a mixture of different photon number terms.

Hence we find that in contrast to previous anaysis for bi-partite systems we find that in general spectral filtering does not improve the purity of the system but introduces mixing in the number degree
of freedom. This is a generic property of filtering and imposes
limitations on the applicability of filtering techniques. The notable
exception to this is in all coincidence-type experiments, where photon
number conservation conditions eliminate mixing in the number degree
of freedom by effectively projecting us back onto a state with known
photon number. For example, present-day in-principle demonstrations of
linear optics quantum computing (LOQC) \cite{bib:KLM01} operate in
coincidence, enabling very narrowband filtering to post-select on
highly overlapping spatio-temporal modes (see, for example,
Refs.~\cite{bib:OBrien03,bib:Pan03,bib:Walther05a}). However, because
these experiments operate in coincidence they are inherently
destructive. Scalable implementations typically require heralded
quantum gates, in which case such filtering is ruled out due to the described mixing
effect. In this context preserving purity of the states is extremely
difficult. This poses one of the major challenges to the
implementation of scalable LOQC.

Next consider the case of finite bandwidth photo-detection. A finite
bandwidth photo-detector can be modeled by considering an ideal
detector preceded by an appropriate spectral filter. Thus, the state
shown in Eq.~\ref{eq:rho_filter} represents the state observed by the
ideal detector. The photon number statistics at the detector will
simply be given by
\begin{equation}
  P_m = \binom{n}{m}{\lambda_1}^{2m} \lambda_0^{2(n-m)},
\end{equation}
compared to the ideal case where $P_m=\delta_{m,n}$. See Ref.~\cite{bib:RohdeRalph06b} for a different treatment of finite bandwidth photo-detection.

\subsection{Homodyne detection}
Next we consider homodyne detection in an effective single-mode description which imposes mode-mismatch if we are not able or allowed to chose the local oscillator mode as underlying spatio-spectral mode. Experimentally we encounter this situation if we consider a quantum network with several -- possibly correlated -- signal modes which exhibit different spatio-spectral characteristics. In such systems
 we have to decide on a reference mode for all channels in the network. The derivation of homodyne detection in presence of mode mismatch is well-understood (cf. Ref.~\cite{bib:Leonhard05,bib:Banaszek02}), but we include this  example to illustrate how our mode decomposition method relates to known results and to emphasize that spectral-spatial filtering can be hidden in the detection.

Homodyne detection is widely used in quantum optics experiments and forms the basis of Optical Homodyne Tomography (OHT)
\cite{bib:Smithey93}, which is used to tomographically reconstruct
the Wigner function of unknown states. The measurement proceeds by
interacting an unknown state with a coherent probe beam. Here an
\emph{indirect} filtering process takes place. Specifically, only
the component of the incident state overlapping with the coherent
probe mode will contribute to the homodyne statistics. As in the
previous example, this effectively traces out the non-overlapping
components. We now consider this in detail.

For simplicity, let us consider a superposition of Fock states of the form
\begin{equation}
\ket\psi = \sum_n c_n\ket{\psi_n},
\end{equation}
that we wish to measure. We will assume the coherent probe
$\ket\alpha$ is formally a superposition of Fock states, characterized
by mode function $\phi(\omega)$. We begin by reexpressing our state
$\ket\psi$ in the eigenmode occupation number representation, where we
choose one of the eigenmodes to be $\phi(\omega)$ (the others are
irrelevant for the calculation). Thus,
\begin{equation}
  \ket\psi = \sum_n c_n \sum_{i=0}^n \lambda_{n,i}\ket{i}_\phi\ket{n-i}_{\bar\phi},
\end{equation}
and the corresponding density matrix is simply
\begin{eqnarray}
  \hat\rho &=& \ket\psi\bra\psi\nonumber\\
  &=& \sum_{n,i,n',i'} {c_n}^* c_{n'} {\lambda_{n,i}}^* {\lambda_{n',i'}}^* \ket{i}_{\phi}\ket{n-i}_{\bar\phi} \bra{n'-i'}_{\bar\phi}\bra{i'}_\phi.\nonumber\\
\end{eqnarray}
As discussed, the components of the incident state which do not
overlap with the coherent probe do dot contribute to the homodyne
statistics and are therefore effectively traced out. Thus, the
observed density matrix is given by
\begin{eqnarray}
  \hat\rho_\mathrm{obs} &=& \mathrm{tr}_{\bar\phi}(\hat\rho)\nonumber\\
  &=& \sum_{n,i,n',i'} {c_n}^* c_{n'} {\lambda_{n,i}}^* {\lambda_{n',i'}}^* \delta_{n-i,n'-i'} \ket{i}\bra{i'}.
\end{eqnarray}
Consider the behavior of this observed state in the limiting
cases. First consider the ideal case, when the incident state is
perfectly mode-matched to the coherent probe. In this case
$\lambda_{i,j}=\delta_{i,j}$ and it follows that
\begin{equation}
  \hat\rho_\mathrm{obs} = \sum_{n,n'} {c_n}^* c_{n'} \ket{n}\bra{n'} = \ket\psi\bra\psi,
\end{equation}
as expected. In the other extreme the incident state is completely
mismatched from the coherent probe. Now $\lambda_{n,i}=\delta_{i,0}$
and the observed state is given by
\begin{equation}
  \hat\rho_\mathrm{obs} = \sum_{n,n'} {c_n}^* c_{n'} \ket{0}\bra{0} = \ket{0}\bra{0}.
\end{equation}
This is also intuitively expected -- in the limit where the incident
state is completely mismatched from the probe, the detection process
just observes a vacuum state. The intermediate cases may be readily
calculated given knowledge of the $\lambda$'s, which are easily
derived from Eq.~\ref{eq:lambda_def} for known $\psi$.

Notice that in this calculation we did not need to explicitly model the homodyning process. Instead, it is abstracted away by decomposing the state into components which can be treated in the ideal sense, and which can be discarded. This is one of the main advantages of our approach, and highlights the underlying physics inherent in the structure of multi-mode optical fields. If we want to study the properties of multi-mode states independent of their spatio-spectral character in an effective single-mode model we have to adopt this strategy to eliminate unobserved modes.

\subsection{Preparation of Schr\"odinger Kitten states}
We now turn our attention to a more specific example, a single-mode scheme for
preparing Schr\"odinger Kitten states through photon subtraction
from a pulsed single-mode squeezed state. This has been experimentally
demonstrated in Ref.~\cite{bib:Ourjoumtsev06}. In this experiment a
squeezed state is prepared via degenerate parametric
down-conversion. A photon is non-deterministically subtracted from
this state by inserting a low reflectivity beamsplitter and
conditioning on detecting a single photon in the reflected mode.
When this post-selection process succeeds, we ought to have a
photon-subtracted squeezed state in the transmitted mode, which
corresponds to the desired Kitten state. We now examine the effect
of the SDF on the preparation of such states.

For a pulsed system we model the Hamiltonian of the parametric down-conversion process as
\begin{equation} \label{eq:nd_dc_ham}
	\hat{H}(t)=\lambda\int\!\!\int \psi(\omega_1,\omega_2) \hat{a}^\dag(\omega_1) \hat{a}^\dag(\omega_2) \,\mathrm{d}\omega_1\,\mathrm{d}\omega_2 + H.c.,
\end{equation}
where $\psi(\omega_1,\omega_2)$ is determined by the crystal properties, phase matching conditions, and pump pulse \cite{bib:Grice97}. As we show in Appendix~\ref{app:pdc_deriv}, the output state leaving the parametric down-conversion crystal is given by
\begin{eqnarray}
	\ket\psi = \sum_{n=0}^\infty \frac{\lambda^n}{n!}\left[\int\!\!\int\psi(\omega_1,\omega_2) \hat{a}^\dag(\omega_1) \hat{a}^\dag(\omega_2)\,\mathrm{d}\omega_1\,\mathrm{d}\omega_2\right]^n\ket{0}.\nonumber\\
\end{eqnarray}
Rewriting this in terms of an eigenmode decomposition we obtain
\begin{equation}
	\ket\psi = \sum_{n=0}^\infty \frac{\lambda^n}{n!}\left[\sum_{ij} \lambda_{ij}\hat{A}_i^\dag \hat{A}_j^\dag \right]^n \ket{0}.
\end{equation}
Next we pass this state through a low-reflectivity beamsplitter,
\begin{eqnarray}
  \ket\psi &=& \sum_{n=0}^\infty \frac{\lambda^n}{n!} \Big[\sum_{ij} \lambda_{ij} \left(\sqrt{1-\eta}\hat{A}_i^\dag + \sqrt{\eta}\hat{B}_i^\dag\right) \nonumber\\
  &&\times \left(\sqrt{1-\eta}\hat{A}_j^\dag + \sqrt{\eta}\hat{B}_j^\dag\right) \Big]^n \ket{0},
\end{eqnarray}
where $\eta$ is the beamsplitter reflectivity. We then post-select
upon detecting a single photon in the reflected mode $B$. In practise
\cite{bib:Ourjoumtsev06}, very narrowband filtering is employed prior
to photo-detection. Suppose the filter has an ideal top-hat
response. Furthermore, we will assume that our eigenbasis
decomposition is in terms of a basis where the first basis function
$\xi_0$ is the filter response function. In the limit of small
beamsplitter reflectivity, $\eta\ll 1$, the conditioned transmitted
state can be approximated as,
\begin{equation} \label{eq:kitten_general} \ket\psi \approx
  \sum_{n=0}^\infty \frac{\lambda^n}{(n-1)!} \left[\sum_{j}
    \lambda_{0j} \hat{A}_j^\dag \right] \left[\sum_{ij} \lambda_{ij}
    \hat{A}_i^\dag \hat{A}_j^\dag \right]^{n-1} \ket{0}.
\end{equation}
In general this state is not a superposition of Fock states, and
therefore not formally a Kitten state. There are two features of this
state that are of particular interest. First, looking at the second
term, it is evident that the photon pairs created by this operator
are, in general, pairwise spectrally entangled. Second, in general the
photon created by the first term has a different spectral form to the
other photons.

For the preparation of ''real'' Schr\"odinger Kitten states we are interessted in the conditions under which the entanglement between the different components vanishes. From the engineering of spontaneous parametric downconverion we know that spectral entanglement of bi-photon states can be eliminated by a careful design of  the generation process \cite{bib:Grice01}. In this case
 the initial state from the down-conversion
process is separable into identical components (i.e. formally a
superposition of Fock states). Thus $\lambda_{ij}$ becomes
separable, i.e. $\lambda_{ij} = \lambda'_i \lambda'_j$, and it follows
that the photon-subtracted state is of the form
\begin{equation}
  \ket\psi \approx \lambda_0 \sum_{n=0}^\infty \frac{\lambda^n }{(n-1)!} \left[\sum_{j} \lambda_j' \hat{A}_j^\dag \right]^{2n-1} \ket{0},
\end{equation}
where the leading factor of $\lambda_0$ reflects the non-determinism
of the conditioning process. Clearly in this case the state is
formally a superposition of Fock states, as required. Note, that in this second analysis we chose our mode decomposition according to the internal structure of the parametric downconversion states. Then the mode of the photon which we use for post-selection gets decoupled from the signal state and narrow band filtering becomes obsolete.
This provides us with a clear strategy to realize Schr\"odinger kitten states in practise such that the spatio-spectral structure of the states can be ignored and an effective single mode treatment becomes valid.

 However, in
general the down-conversion process does not produce separable photon
pairs, but highly spectrally entangled pairs as a result of energy
conservation conditions. Thus, this technique will not
formally produce Kitten states, but rather a more general class of
states exhibiting the same photon number distribution as Kitten
states. We might be tempted to utilize narrowband filitering for purifiying the state. Though,
recall from the previous sub-sections that spectral filtering is equivalent to a decomposition into one signal and one unobserved eigenmode with a subsequent tracing over the unobserved mode.
 When a photon pair is spectrally entangled it is not possible for both
photons to simultaneously have perfect mode overlap with the defined signal mode   -- an entangled state cannot have unit overlap with a separable one.
 This will degrade the purity of the states with  correlations between $ \hat{A}_i^\dag, \hat{A}_j^\dag$.
 When performing homodyne detection, imperfect mode overlap with the coherent probe beam also leads to a tracing out of the orthogonal components of the measured mode. This effectively introduces mixing with the vacuum state.  Thus, in general we expect this procedure to result in observed mixing with the vacuum state, even in the presence of perfect mode-matching, i.e. equivalent spectrum for sigal and local oscillator mode.
 Note that in the experiment of Ref.~\cite{bib:Ourjoumtsev06} significant mixing of this type was observed, ranging from $29\%-36\%$. Based on our results we expect that this arises from a combination of mode-mismatch and spectral entanglement effects. It is important to recognize that the later effect is, in general, inherent to this state preparation technique. This type of mixing cannot be compensated for through filtering, post-selection or improved alignment and should be distinguished from experimental imperfections.

In conclusion the fideltiy for the conditional preparation of Schr\"odinger kitten states is already constrained by the generation of the squeezed states.  Spectral correlations between differnt photon numbers degrade the achievable purity such
that the best avenue for improving state fidelity is to engineer superpostions of true Fock states in the first instance.

\subsection{Conditional preparation of Fock states via non-degenerate parametric down-conversion}
Finally we consider the example of conditional state preparation via
non-degenerate parametric down-conversion. This technique is widely
used in the preparation of single photon states for quantum optics
applications, including quantum information processing ones. We
consider the more general case where we postselect upon detecting $n$
photons in one arm of a non-degenerate down-converter and examine the
form of the conditional state in the other arm. In the non-degenerate
case the Hamiltonian differs only slightly from
Eq.~\ref{eq:nd_dc_ham}, and takes the form,
\begin{equation}
  \hat{H}(t)=\lambda\int\!\!\int \psi(\omega_1,\omega_2) \hat{a}^\dag(\omega_1) \hat{b}^\dag(\omega_2) \,\mathrm{d}\omega_1\,\mathrm{d}\omega_2 + H.c.
\end{equation}
Following a similar derivation as per the previous example, the
prepared state is of the form
\begin{equation} \label{eq:nd_dc_output} \ket\psi = \sum_{n=0}^\infty
  \frac{\lambda^n}{n!}\left[\sum_{ij} \lambda_{ij} \hat{A}_i^\dag
    \hat{B}_j^\dag \right]^n \ket{0}.
\end{equation}
Now suppose we post-select upon detecting $m$ photons in mode $A$. In
general the detection process may involve filtering. We will use the
usual trick of choosing our eigenmode basis to contain the detector
mode as one of its elements, $\xi_0$. Thus, the detector effectively
applies the projector $(\ket{\xi_0}\bra{\xi_0})^m$. Upon applying this
projector, the conditionally prepared state is
\begin{equation}
  \ket{\psi_\mathrm{cond}} \approx \frac{\lambda^m}{m!}\left[\sum_j \lambda_{0j} \hat{B}_j^\dag \right]^m \ket{0},
\end{equation}
where we ignore terms $n>m$. This approximation is valid as long as $\lambda \ll 1$, which is typically the case if the downconversion source is operated in the low gain regime.
 We can relabel the term in the bracket to obtain
\begin{equation} \label{eq:cond_prep}
\ket{\psi_\mathrm{cond}} \approx \frac{\lambda^m}{m!}\left(\tilde{B}^\dag\right)^m \ket{0},
\end{equation}
where $\tilde{B}^\dag = \sum_j \lambda_{0j} \hat{B}_j^\dag$ is the
mode creation operator characterized by the marginal distribution of
the joint SDF over mode $B$. Importantly, in contrast to the previous example the
conditionally prepared state in Eq.~\ref{eq:cond_prep} is always
formally an $m$-photon Fock state, independent of $\vec\lambda$ making this approach to state preparation very suitable for quantum
optics applications. Note that this feature arises from the
particular form of Eq.~\ref{eq:nd_dc_output}, specifically the
pairwise spectral entanglement, whereas for a completely arbitrary
multi-photon state this property will not apply. This highlights the
variability in the behavior of systems under filtering and
measurement, and the need to analyze them on an individual basis. In
some cases, as per the previous examples, filtering has a major impact on the form of the final state of the system, while in others, as
here, it does not.

In this analysis we have assumed the detector projects onto a particular, well-defined mode $\ket{\xi_0}$. As mentioned in the introduction realistic detectors with single photon sensitivity, namely avalanche photo diodes, lack the time resolution to distinguish the temporal modes. This leads to the annihilation of off-diagonal frequency modes \cite{bib:URen05}. For this reason narrowband filtering is an effective tool in increasing the purity of conditionally prepared states, since it effectively forces the detector to respond to only a single well-defined mode \cite{bib:RohdeRalph06b}. Our result of this subsections clarify that the the strategy to improve the purity by narrow band filtering, which is known for single photon state, can be extended to Fock states of higher order number states if no superpostions of different photon numbers are considered.

\section{Conclusion} \label{sec:conclusion}
We have discussed the spectral structure and decomposition of multi-photon states. We considered the general properties of the spectral distribution function and introduced an approach for decomposing it into discretized orthogonal `eigenmodes'. This leads naturally to an occupation number representation for general optical states. The advantage of this representation is that it simplifies many types of calculations. There are two main reasons for this. First, we have great freedom in choosing the basis for the decomposition. So, for example, when considering the interaction between two states, we can decompose one state into a basis consisting of components overlapping with and orthogonal to the other state. This allows for very straightforward treatment. Second, because the discrete eigenmodes are orthogonal, many types of evolution (e.g. all linear optical interactions) can be calculated by acting the relevant
evolution independently on each of the eigenmodes. This allows us to adopt an effective single-mode representation where we choose one underlying spatio-spectral reference mode and define a photon number respresentation corresponding to this mode. We have shown that homodyne detection can be
understood in this picture. Here mode mismatch is to be interpreted not only as an experimental imperfection but as a limitation inherent to a system comprising modes with correlated or different spatio-spectral characteristics. We have applied these techniques to several other examples, which provides further
 insight into the impact of the internal structure of multimode states on quantum state manipulation. For the Hong-Ou-Mandel interference our methods highlight very clearly that the symmetry of states defines the visibility of the interference. Our analysis of spectral filtering as well as the Schr\"odinger Kitten state preparation proves that spectral filtering in general does not purify quantum states. The detection of vacuum mixing for conditionally prepared Schr\"odinger Kitten states, which is commonly attributed solely to experimental imperfections, also arises from the spatio-spectral correlation of the internal structure of states. To eliminate these effects the best strategy is to engineer pure squeezed states which do not exhibit spatio-spectral correlations. In contrast to this example the generation of Fock states of higher photon number can be accomplished by narrow band filtering of a parametric downconversion state.

In conclusion, our examples illustrate that different optical systems have to be analyzed on an individual basis. In many situations an adapted mode decomposition simplifies calculations and provides the means to identify the impact of spatio-spectral structure. We believe the presented techniques will prove useful in the theoretical analysis of quantum optical systems, particularly in situations where `non-ideal' effects such as mode-mismatch, photon distinguishability and imperfect detectors are considered.

\appendix

\section{Proof that $\mathcal{N}=n!$ if and only if $\psi(.)$ exhibits full permutation symmetry} \label{app:n_fact_proof}
By definition,
\begin{equation}
\mathcal{N} = \int \psi(\omega_1,\dots,\omega_n)^* \sum_{P\in S_n}
\psi(P[\omega_1,\dots,\omega_n]) \,\mathrm{d}\vec\omega.
\end{equation}
We will adopt the shorthand $\psi = \psi(\omega_1,\dots,\omega_n)$ and $\psi_P = \psi(P[\omega_1,\dots,\omega_n])$. We now employ the Cauchy-Schwarz inequality,
\begin{equation}
|\langle \psi,\psi_P\rangle| \leq ||\psi||.||\psi_P||,
\end{equation}
where $||\psi|| = \langle \psi,\psi\rangle$. In our case we have $||\psi||=||\psi_P||=1$. Thus,
\begin{equation}
|\langle \psi,\psi_P\rangle| \leq 1,
\end{equation}
with equality for $\psi=\psi_P$. The expansion for $\mathcal{N}$ contains $n!$ terms, each of the form $\langle \psi,\psi_P\rangle$ which have magnitude less than or equal to unity. Furthermore, since $\mathcal{N}=\langle\psi|\psi\rangle$, $\mathcal{N}\in \mathbb{R}$. In general $\langle \psi,\psi_P\rangle \in \mathbb{C}$, but if
$\mathcal{N}=n!$ then necessarily $\langle \psi,\psi_P\rangle = 1\,\,\forall\,\,P$, i.e $\psi=\psi_P\,\,\forall\,\,P$.

\section{Proof that a basis containing an arbitrary integrable function as one of its elements always exists} \label{app:function_decomp_proof}
Let $\mathcal{H}$ be a separable Hilbert space with countable infinite basis $X = \{x_{n}\}$. Let $f$ be a normalized function $\mathcal{H}\rightarrow\mathcal{H}$. We wish to show that the set $X' = X \backslash x_k \cup f$ is a linearly independent set which can be used to create the orthonormal basis $Y = \{y_n\}$ with $y_0 = f$.

Since $f\in\mathcal{H}$ and $X$ forms a basis for $\mathcal{H}$ there must exist at least one $x_k$ such that $\langle x_k , f \rangle \neq 0$. Then $X \backslash x_k$ remains linearly independent but is no longer complete. Adding $f$ to $X \backslash x_k$ restores completeness, but $X' = X \backslash x_k \cup f$ is not an orthonormal basis because there exist $i\neq k$ such that $\langle x_i , f \rangle \neq 0$. Because $X$ is countable, the number of expansion coefficients for $f$ is also countable, so the expansion of any $g \in \mathcal{H}$ is possible with a finite number of expansion coefficients using $X'$. The integral form of H\"older's inequality (see, for example, Ref.~\cite{bib:Cheney01}) guarantees the existence of the scalar product defined over $X'$. Since $X'$ is linearly independent, an orthonormal basis $Y$ can be constructed using the Gram-Schmidt orthogonalization procedure. Note that this argument relies on the separability of the underlying Hilbert space. This is fulfilled for most systems of interest, but there are counterexamples (the standard one being an infinite spin chain) which do not fulfill this requirement. In such cases, convergence issues will need to be investigated separately.

\section{Four-photon normalization factor in four-photon interference}
\label{app:N_4_proof}
We wish to calculate the normalization factor $\mathcal{N}_4$ for a state of the form
\begin{equation}
\ket{\psi_4} = \hat{A}^\dag_{\varphi_1} \hat{A}^\dag_{\varphi_1} \hat{A}^\dag_{\varphi_2} \hat{A}^\dag_{\varphi_2} \ket{0}.
\end{equation}
We have
\begin{equation}
\mathcal{N}_4 = \langle\psi_4|\psi_4\rangle = \bra{0} \hat{A}_{\varphi_1} \hat{A}_{\varphi_1} \hat{A}_{\varphi_2} \hat{A}_{\varphi_2} \hat{A}^\dag_{\varphi_1} \hat{A}^\dag_{\varphi_1} \hat{A}^\dag_{\varphi_2} \hat{A}^\dag_{\varphi_2} \ket{0}.
\end{equation}
Summing over all combinations in which terms from the left hand side of the expansion can act on terms from the right hand side we obtain
\begin{equation}
\mathcal{N}_4 = 4{\gamma_{11}}^2{\gamma_{22}}^2 + 4\gamma_{11}\gamma_{22}\gamma_{12}\gamma_{21} + 4{\gamma_{12}}^2{\gamma_{21}}^2,
\end{equation}
where
\begin{equation}
\gamma_{ij} = \bra{0} \hat{A}_{\varphi_i} \hat{A}_{\varphi_j}^\dag \ket{0}.
\end{equation}
Note that $\gamma_{11}=\gamma_{22}=1$. Thus,
\begin{equation}
\mathcal{N}_4 = 4(1+\gamma+\gamma^2),
\end{equation}
where
\begin{equation}
\gamma = \gamma_{12}\gamma_{21} = \left| \int \varphi_1(\omega)^* \varphi_2(\omega) \,\mathrm{d}\omega \right|^2.
\end{equation}

\section{Derivation of the PDC Hamiltonian}\label{app:pdc_deriv}
Following \cite{bib:Perina06}, the time-dependent Hamiltonian for the PDC
process is given by
\begin{equation}
\hat{H}(t) \propto \sum_{k_{\text{s}}}\sum_{k_{\text{i}}}
  \tilde{\psi}(k_{\text{s}}, k_{\text{i}},t)
    \hat{a}^{\dagger}_{\text{s}}(k_{\text{s}})
    \hat{a}^{\dagger}_{\text{i}}(k_{\text{i}}) + \text{h.c.}
\end{equation}
where h.c.~denotes hermitian conjugation and \(\psi(k_{\text{s}},
k_{\text{i}},t)\) characterizes the joint spatio/temporal structure of
the signal and idler photons. Collecting all creation operators in
\(\hat{H}^{(-)}(t)\) and all annihilation operators in
\(\hat{H}^{(+)}(t)\), we can write the Hamiltonian as
\begin{equation}
  \hat{H}(t) = \hat{H}^{(-)}(t) + \hat{H}^{(+)}(t),.
\end{equation}

Time propagation of the output state is governed by the Schr\"odinger
equation. We obtain a solution to the differential equation using the
standard integral ansatz which can be found in any standard textbook
on quantum mechanics (e.g., \cite{bib:Merzbacher98}). This leads to state of
the form \(\ket{\psi} = \sum_{n} \ket{\psi_{n}} =
\sum_{n}\hat{H}_{n}\ket{0}\) where
\begin{align}
  \ket{\psi_{0}} =& \ket{0}\\
  \ket{\psi_{n}} =& \left(\frac{-i}{\hbar}\right)^{n}
    \int_{-\infty}^{\infty}\text{d}\tau_{1}
    \int_{-\infty}^{\tau_{1}}\text{d}\tau_{2}\cdots
    \int_{-\infty}^{\tau_{n-1}}\text{d}\tau_{n}\nonumber\\
    &\times
\hat{H}(\tau_{1})\cdots\hat{H}(\tau_{n})\ket{0}.\label{pdc_prop}
\end{align}

Consider the product of Hamiltonians for different times \(\tau_{k}\):
\begin{align}
  \prod_{k=1}^{n}\hat{H}(\tau_{k}) &=
  \prod_{k=1}^{n}(\hat{H}^{(-)}(\tau_{k}) +
  \hat{H}^{(+)}(\tau_{k})) \\
  & = \sum_{S\in(\pm)^{*}}\prod_{k=1}^{n}\hat{H}^{(S[k])}(\tau_{k}).
\end{align}

Assuming that the number of generated photons is negligible compared
to the number of possible field modes, annihilation operators
contained in \(\hat{H}^{(+)}\) will predominantly hit a vacuum mode.
Thus \(\prod_{k=1}^{n}\hat{H}^{(S[k])}(\tau_{k})\ket{0} = 0\) if there
is at least one \(k\) such that \(S[k] = \) ``+''. Only contributions
with \(S[k] = \) ``-'' for all \(k\) will effectively contribute. This
simplifies Eqn.~\ref{pdc_prop} as follows.
\begin{align}
  \hat{H}_{n} \approx& \left(\frac{-i}{\hbar}\right)^{n}
  \int_{-\infty}^{\infty}\text{d}\tau_{1}
  \int_{-\infty}^{\tau_{1}}\text{d}\tau_{2}\cdots
  \int_{-\infty}^{\tau_{n-1}}\text{d}\tau_{n}\nonumber\\
  &\times \hat{H}^{(-)}(\tau_{1})\cdots\hat{H}^{(-)}(\tau_{n})\\
  =& \left(\frac{-i}{\hbar}\right)^{n}\frac{1}{n!}
  \left(\int_{-\infty}^{\infty}\text{d}\tau
    \hat{H}^{(-)}(\tau)\right)^{n}\\
  =&
  \sum_{n=0}^{\infty}\frac{\lambda^{n}}{n!}\left(\iint_{-\infty}^{\infty}
    \text{d}\omega_{1}\text{d}\omega_{2}\psi(\omega_{1},
    \omega_{2})\hat{a}^{\dagger}(\omega_{1})\hat{a}^{\dagger}(\omega_{2})
  \right)^{n}\label{pdc_prop_s}
\end{align}

This is the form of the Hamiltonian conventionally employed for the
analysis of parametric downconversion which we also employ in this
paper.

\begin{acknowledgments}
We thank Malte Avenhaus, Andreas Eckstein, Tobias Moroder, Ian A. Walmsley, Timothy Ralph and Martin Plenio for helpful discussions. PR acknowledges support from the Australian Research Council, Queensland State Government, and DTO-funded U.S. Army Research Office Contract No. W911NF-05-0397.
\end{acknowledgments}

\bibliography{paper}

\end{document}